\documentstyle [12pt]{article}
\advance\voffset by -1 in
\advance\hoffset by -.5 in
\def\f{\phi}
\def\l{\lambda}
\def\m{\mu}

\def\p{\partial}

\def\O{\Omega}
\def\Om0{\Omega^0}
\def\O1{\Omega^1}

\def\hw{\hat{w}}

\def\res{{\rm res}\:}

\def\Ga{\Gamma}
\begin{document}
\centerline{{\bf On the constrained KP hierarchy}}

\vspace{.2in}
\centerline{{\bf L.A.Dickey}}

\vspace{.1in}
\centerline{Dept. Math., University of Oklahoma, Norman, OK 73019\footnote{
Internet: "ldickey@nsfuvax.math.uoknor.edu"}}

\vspace{.3in}
\centerline{{\bf Abstract}}

\vspace{.2in}
\small
\parbox{5.5in}{An explanation for the so-called constrained hierarhies is
presented by linking them with the symmetries of the KP hierarchy. While
the existence of ordinary symmetries (belonging to the hierarchy) allows one to
reduce the KP hierarchy to the KdV hierarchies, the existence of additional
symmetries allows to reduce KP to the constrained KP.}
\normalsize

\vspace{.2in}
There are several papers published lately concerning the so-called
``constrained KP hierarchies", [1-4]. Closely related to this topic is the
``two-boson representation" of the KP hierarchy also given in a series of
papers [5-7].

In this note we are going to explain, in a not very formal way, our viewpoint
on the origin of the special form of those constraints linking them with
well-known subjects such as symmetries of the KP hierarchy.

Roughly speaking, in the same way as the existence of ordinary symmetries
(belonging to the hierarchy) allows one to reduce the KP hierarchy to the KdV
hierarchies, the existence of additional symmetries allows to reduce KP to
the ``constrained KP". More than that, the possibility of this type of
constraints is equivalent to the existence of the additional symmetries.

The proof that the above mentioned constraints are compatible with the
hierarchy is rather simple, see quoted papers or the last paragraph of this
note, while our main goal is to put the problem into a wider context giving an
interpretation of the constrained hierarchy from the point of view of the
symmetries. It is based on a new formula for the generator of additional
symmetries [8]. For convenience, since that paper is not published yet, the
proof of the formula is also enclosed into this note as an Appendix.

We are not discussing Hamiltonian properties of the equations; some of the
above cited works were devoted to their study.

The constrained hierarchy is the KP hierarchy restricted to pseudo-differential
operators of the form $$L^n=\p^n+U_{n-2}\p^{n-2}+...+U_0+w\p^{-1}w^*,~~\p=d/dx
$$ where $w$ and $w^*$ satisfy the equations $$\p_mw=L_+^mw,~~\p_mw^*=-L_+^{m*}
w^*$$ and $L_+^{m*}$ is the operator adjoint to $L_+^{m}$. The equations are
exactly the same as those for the Baker and the adjoint Baker functions.
However, $w$ and $w^*$ are not supposed to be necessarily the Baker and the
adjoint functions (i.e., eigenfunctions of the operators $L$ and $L^*$), just
any solutions.\\

Recall some well-known definitions and results. Let $$L=\p+u_1\p^{-1}+u_2\p^{-
2}+...$$ be a pseudo-differential operator ($\Psi$DO), the coefficients $u_k$
being taken as independent generators of a differential algebra. The
KP-hierarchy is the set of equations $$\p_mL=[L_+^m,L],~~\p_m=\p/\p t_m
\eqno{(1)}$$ where $t_m$, $m=1,2,...$ are some variables, and the subscript
+ symbolizes preserving only terms with non-negative powers of $\p$ (similarly,
the subscript $-$ below for the negative part). All the equations
commute and can be solved simultaneously. It follows from (1) that $\p_mL^n
=[L_+^m,L^n]$ and $$\p_mL_-^n=[L_+^m,L^n]_-=[L_+^m,L_-^n]_-.$$This implies that
if $L_-^n$ initially vanishes then it remains zero along the trajectory. This
allows to restrict the hierarchy to the operators $L$ such that $L_-^n$ is
identically zero, i.e., $L^n$ is a differential operator. We call this the
$n$-restricted KP or the $n$th KdV-hierarchy.

In many cases it is convenient to represent the KP operator $L$
in a ``formal dressing" form: $$L=\f \p\f^{-1}$$ where
$\f$ is a $\Psi$DO $\f=\sum_0^\infty w_i\p^{-i}$ with $w_0=1$. This yields
expressions of $u_i$ as differential polynomials in terms of $w_i$.
The dressing operator $\f$ is determined up to a multiplication on the right
by a series in $\p^{-1}$ with constant coefficients starting with $1$. In terms
of $\f$, the equations of the hierarchy are $$ \p_m\f=-L_-^m\f.$$

Let $\xi(t,z)=\sum_1^\infty t_kz^k$. Then
$$w(t,z)=\f\exp\xi(t,z)=\sum_0^\infty
w_iz^{-i}\exp\xi(t,z)=\hw(t,z)\exp\xi(t,z)
$$ is called the (formal) {\em Baker, or wave, function}. The Baker function
satisfies equations $$Lw=zw,~~\p_mw=L_+^mw.$$

Let $\f^*$ be the formal conjugate to $\f$ (by definition, $(f\p)^*=-\p\circ f
$). The function $w^*(t,z)=(\f^*)^{-1}\exp(-\xi(t,z))=\hw^*(t,z)\exp(-\xi(t,z))
$ is called the {\em adjoint Baker function}. The adjoint Baker function
satisfies the equations $$ L^*w^*=zw^*,~~\p_mw^*=-L_+^{m*}w^*.$$

Nothing prevents introducing constraints more general than $L_-^n=0$, namely,
linear combinations of them, $$L_-^n-\sum a_lL_-^l=0,$$ which is a very natural
generalization (of course, the term $L_-^n$ can be included into the linear
combination, however, it is more convenient to write the constraint as above).
The proof of compatibility of those constraints with the
hierarchy remains the same. In particular, we can take, as $\sum a_lL_-^l$, a
generating function of equations of the hierarchy, $T(\l)=\sum_{-\infty}^\infty
L_-^l\l^{-l-1}$, the so-called resolvent (see [11]). It can be proven ([11],
(7.6.2)) that $T(\l)=w(t,\l)\p^{-1}w^*(t,\l)$. Thus, the hierarchy can be
restricted to operators $L$ such that $$L^n=L_+^n+w(t,\l)\p^{-1}w^*(t,\l).\eqno
{(2)}$$ Note also the following useful
representation of the resolvent ([11], (7.6.6), (7.6.3))
$$T(\l)=(\p-\chi)^{-1}S
=\sum_0^\infty \p^{-1-k}P_k(\chi)S\eqno{(3)}$$ where $\chi=w'/w$ and $S=ww^*$.
They have an advantage being, in contrast to $w$ and $w^*$, local expressions
in terms of coefficients of the operator $L$; $P_k$ are the Fa\`a di Bruno
differential polynomials.

Those formulas do not make much sense yet since $w$ and $w^*$ are formal series
in $\l$. However, $w$ and $w^*$ can be constructed as actual functions on a
Riemann surface while the series appear only as their asymptotic expansions
in a neighborhood of an essential singular point. Then $w$ and $w^*$ in
Eq.(2) are values of these functions at a particular point $\l$. The equations
$Lw=\l w$ and $L^*w^*=-\l w^*$ become meaningless since the action of a
$\Psi$DO $L$ on functions is not defined. The equations
$$\p_mw=L_+^mw,~~\p_mw^*
=-L_+^{m*}w^*,\eqno{(4)}$$ this is all what remains as a definition of $w$ and
$w^*$. Can we understand $w$ and $w^*$ in (2) as arbitrary solutions of (4)
just disregarding the argument $\l$? The proof in the last paragraph will give
a positive answer to this question. However, the preceding analysis cannot
explain this fact. Of course, linear combinations of all $w(t,\l)$ with
various $\l$ give spectral representation of all solutions of the equation $\p_
mw=L_+^mw$  (no one has ever proven that every solution of the initial
value problem can be represented as a series in eigenfunctions, though it is a
plausible conjecture; this is why we call this reasoning not very formal).
However, we cannot use this in (2) because the last term is
quadratic. To overcome this difficulty, we need a bilinear expression $w(t,\m)
\p^{-1}w^*(t,\l)$ in (2), rather than quadratic $w(t,\l)\p^{-1}w^*(t,\l)$. And
this is precisely the generator of the so-called additional symmetries, what we
are going to discuss next.\\

Commutativity of flows generated by the equations of the KP hierarchy
means that each of them is a symmetry for all the others. There are, however,
symmetries which do not belong to the hierarchy itself. They are called
additional symmetries. They do not commute between themselves. The hallmark
of these symmetries is their explicit dependence on the variables $t_i$.
We use additional symmetries in a form given them by Orlov and Schulman [10].

Dressing an obvious relation $[\p_k-\p^k,\p]=0$ we obtain the equation
of the hierarchy $[\p_k-L_+^k,L]=0$.
There is another operator commuting with $\p_k-\p^k$. This is $$\Gamma=
\sum_1^\infty t_ii\p^{i-1}.$$ Dressing the relation $[\p_k-\p^k,\Gamma]=0$ one
obtains $[\p_k-L_+^k,M]=0$, or $$ \p_kM=[L_+^k,M],~{\rm where}~M=\f\Gamma\f^{
-1}.\eqno{(5)}$$

{}From (5) it follows that $$\p_k(L^lM^m)=[L_+^k,L^lM^m].\eqno{(6)}$$

An additional symmetry is a differential equation
$$\p^*_{lm}\f=-(M^mL^l)_-\f$$ where $\p^*_{lm}$ symbolizes a derivative
with respect to some additional variable $t^*_{lm}$. In terms of the operator
$L$ this definition becomes $$\p^*_{lm}L=-[(M^mL^l)_-,L].$$ The operators $\p^*
_{lm}$ commute with all $\p_k$, i.e., they determine symmetries, indeed.

One can introduce a generating function of these symmetries
$$Y(\l,\m)=\sum_{m=0}^\infty{(\m-\l)^m\over m!}\sum_{l=-\infty}^\infty \l^
{-l-m-1}(M^mL^{m+l})_-.\eqno{(7)}$$  The following theorem holds ([8], see
also the Appendix to this note).\\

{\bf Theorem.} {\sl The operator $Y(\l,\m)$ is equal to} $$Y(\l,\m)=
w(t,\m)\p^{-1}w^*(t,\l).\eqno{(8)}$$
For $\l=\m$ we obtain resolvents.

Let us consider the following constraint: $$L_-^n=(M^mL^{m+l})_-.
$$ We have $$\p_k(L_-^n-(M^mL^{m+l})_-)=[L_+^k,L^n-M^mL^{m+l}]_-=
[L_+^k,(L^n-M^mL^{m+l})_-]_-.$$ Hence, if $(L^n-M^mL^{m+l})_-$ is zero at the
initial moment, it remains zero along the whole trajectory, and the constraint
$L_-^n=(M^mL^{m+l})_-$ is compatible with the hierarchy. Actually, what we have
done, we restricted the hierarchy to operators $$L^n=L_+^n+(M^mL^{m+l})_-.
\eqno{(9)}$$ Moreover, $(M^mL^{m+l})_-$ can be replaced by
any linear combination of such expressions with different $m$ and $l$. The same
proof remains valid. Our claim is that in this manner the constrained hierarchy
can be obtained.

One can take the generating function $Y(\l,\m)$ as a particular case of a
linear combination. According to the above theorem this is $w(t,\m)\p^{-1}w^*(t
,\l)$.

Thus, we obtain a restriction to the operators of the form $$L^n=L_+^n+w(t,\m)
\p^{-1}w^*(t,\l).\eqno{(10)}$$ If the Baker and the adjoint functions exist as
genuine functions of $\l$ and $\m$, then one can take their values at fixed
points. Now, $w(t,\m)$ can be replaced by any linear combination of them with
different $\m$, the same with $w^*$. This allows to consider restrictions to
operators $$L^n=L_+^n+w(t)\p^{-1}w^*(t)\eqno{(11)}$$ where $w(t)$ and $w^*(t)$
are solutions to Eq.(4) with arbitrary initial conditions.

The above discussion was not a rigorous proof. The proof of possibility of
restriction (11) is much simpler, see below.
Our goal was to show that the appearance of the expression $w(t)\p^{-1}
w^*(t)$ is not accidental, it is a consequence of deep connections with
the symmetries.\\

The proof follows from the equation
$$\p_k(w(t,\m)\p^{-1}w^*(t,\l))=[L_+^k,w(t,\m)\p^{-1}w^*(t,\l)]_-$$ which is
easy to verify. Indeed, for an arbitrary function $f$, $L_+^{k}f=L_+^{k}
\circ f+A\p$ where $A$ is a differential operator. The obvious equality
$f\p=\p\circ f+(-\p f)$ implies that $L_+^{k*}f=fL_+^k+\p\circ B$ where $B$ is
another differential operator. Now, $$\p_k(w(t,\m)\p^{-1}w^*(t,\l))=
\p_k(w(t,\m)\p^{-1}w^*(t,\l))_-$$ $$=(L_+^kw(t,\m)\p^{-1}w^*(t,\l)-w(t,\m)\p^
{-1}L_+^{k*}w^*(t,\l))_-$$ $$=(L_+^k\circ
w(t,\m)\p^{-1}w^*(t,\l)-w(t,\m)\p^{-1}
w^*(t,\l)L_+^k)_-=[L_+^k,w(t,\m)\p^{-1}w^*(t,\l)]_-.$$ The terms with
differential operators $A$ and $B$ vanish since they do not contribute to the
negative part of the whole expression, and the subscript $``-"$ kills them.
The proof can be accomplished as we did this above.
Relation of thus constrained KP to the so-called ``two-boson representation"
one can find in an article by Depireux and Schiff [5], Eq.(2.10). Apparently,
it must be connected with the representation of the resolvent in the form (3).
\\

{\bf Appendix.} {\sl Proof of Eq.(8).}\\

{\bf Lemma.}  {\sl Let $P$ and $Q$ be two $\Psi$DO, then
$$\res_z[(Pe^{xz})\cdot(Qe^{-xz})]=\res_\p PQ^*$$
where $Q^*$ is the formal adjoint to $Q$.}\\

The proof is in a straightforward verification.

It is quite obvious that for every $\Psi$DO $P$ the equality $P_-=
\sum^\infty_1\p^{-i}\res_\p\p^{i-1}P$ holds. We have
$$(M^mL^{m+l})_-=(\phi\Ga^m\p^{m+l}\phi^{-1})_-
=\sum^\infty_1\p^{-i}\res_\p\p^{i-1}\phi\Ga^m\p^{m+l}\phi^{-1}\,.$$
According to Lemma this can be written as
$$(M^mL^{m+l})_-=\sum^\infty_1\p^{-i}\res_z\p^{i-1}\phi
\Ga^m\p^{m+l}e^{\xi(t,z)}(\phi^*)^{-1}e^{-\xi(t,z)}\,.$$
Taking into account that
$$\Ga\exp\xi(t,z)=\sum^\infty_1t_ii\p^{i-1}\exp\xi(t,z)=\sum^\infty_1t_i
iz^{i-1} \exp\xi(t,z)=\p_z\exp\xi(t,z)$$
and that $\phi$ commutes with $\p_z$ we have
$$(M^mL^{m+1})_-=\res_z\sum^\infty_1\p^{-i}(z^{m+l}\p^m_zw)^{(i-1)}\cdot
w^*=\res_z z^{m+l}\p^m_z w\cdot\p^{-1}\cdot w^*\,.$$
Now,
$$Y(\l,\mu)=\res_z\sum^\infty_{m=0}\sum^\infty_{l=-\infty}
\frac{z^{m+l}}{\l^{m+l+1}}\cdot\frac 1{m!}(\mu-\l)^m\p^m_zw\cdot\p^{-1}\cdot
w^*.$$It is easy to see that if $f(z)$ is a Laurent series in $z$ then res$_z
\sum _{l=-\infty}^\infty z^{m+l}\l^{-m-l-1}f(z)=f(\l),$ and
$$Y(\l,\mu)=\exp((\mu-\l)\p_\l)w(t,\l)\cdot\p^{-1}\cdot
w^*(t,\l)=w(t,\mu)\cdot\p^{-1}\cdot w^*(t,\l).~\Box $$

{\bf References.}\\

\noindent 1. Orlov A. Symmetries for unifying different soliton systems into
a single hierarchy, Preprint IINS/Oce-04/03, Moscow, 1991\\

\noindent 2. Bing Xu, A unified approach to recursion operators of the reduced
1+1-dimensional systems, Preprint, Hefei, 1992\\

\noindent 3. Cheng Y., J. Math. Phys. 33, 3774, 1992.\\

\noindent 4. Oevel W. and Strampp W., Constrained KP hierarchy and
bi-Hamiltonian structures, Com. Math. Phys., 157, 51-81, 1993.\\

\noindent 5. Depireux D.A., and Schiff J., On the Hamiltonian structures and
the reductions of the KP hierarchy, preprint LASSNS-HEP-92/66, 1992. \\

\noindent 6. Aratyn H., Ferreira L.A., Gomes J.F., and Zimerman A.H., On
two-current realization of KP hierarchy, 1993, to appear in Nucl. Phys. B.\\

\noindent 7. Bonora L., and Xiong C.S., Multi-field representations of the KP
hierarchy and multi-matrix models, preprint SISSA-ISSAS 57/93/EP, hep-th
9305005, 1993.\\

\noindent 8. Dickey L.A., On additional symmetries of the KP hierarchy and
Sato's B\"acklund transformation, to be published in CMP, 1994.\\

\noindent 9. Adler M., Shiota T., and van Moerbeke P., From the
$w_\infty$-algebra to its central extension: a $\tau$-function approach, 1993,
to appear.\\

\noindent 10. Orlov A. Yu. and Shulman E. I., Additional symmetries for
integrable and conformal algebra representation,
Lett. Math. Phys., 12, 171, 1986.\\

\noindent 11. Dickey, L.A., Soliton equations and integrable systems,
Advanced Series in Math. Phys., vol. 12, World Scientific, Singapore, 1991.\\

\end{document}